\documentclass[11pt]{article}
\usepackage[a4paper,margin=2.2cm]{geometry}
\usepackage[T1]{fontenc}
\usepackage[utf8]{inputenc}
\usepackage{lmodern}
\usepackage{booktabs,tabularx,longtable,array,multirow}
\newcolumntype{L}{>{\raggedright\arraybackslash}X}
\usepackage{graphicx}
\usepackage{float}
\usepackage{caption}
\usepackage{amsmath,amssymb}
\usepackage{xcolor}
\definecolor{DeepBlue}{HTML}{183A59}
\definecolor{Slate}{HTML}{4A5568}
\definecolor{SoftGray}{HTML}{F3F6F8}
\usepackage{hyperref}
\hypersetup{colorlinks=true,linkcolor=DeepBlue,urlcolor=DeepBlue,citecolor=DeepBlue}
\usepackage{fancyhdr}
\usepackage{titlesec}
\usepackage{setspace}
\usepackage{enumitem}
\titleformat{\section}{\large\bfseries\color{DeepBlue}}{\thesection.}{0.4em}{}
\begin{document}
\begin{titlepage}
\centering
{\Huge\bfseries Bolivia and an IMF Extended Fund Facility:\\[0.25em] Financial Sustainability, Verifiable Social Sustainability, and Net Financing Additionality\par}
\vspace{1.1cm}
{\Large Ricardo Alonzo Fern\'andez Salguero\par}
\vspace{0.2cm}
{\normalsize ORCID: \href{https://orcid.org/0000-0002-4189-961X}{0000-0002-4189-961X}\par}
\vspace{0.2cm}
{\normalsize DOI: \href{https://doi.org/10.5281/zenodo.21232618}{10.5281/zenodo.21232618}\par}
\vfill
\begin{minipage}{0.9\textwidth}
\small
\textbf{How to cite.} Fern\'andez Salguero, Ricardo Alonzo. \textit{Bolivia and an IMF Extended Fund Facility: Financial Sustainability, Verifiable Social Sustainability, and Net Financing Additionality}. DOI: 10.5281/zenodo.21232618.
\end{minipage}
\vfill
{\large July 2026\par}
\end{titlepage}
\begin{abstract}
This paper evaluates stabilization scenarios for Bolivia under external-financing stress by jointly modeling financial sustainability, verifiable social sustainability and the net additionality of multilateral financing. The main correction is that financing already received, contracted or expected from multilateral institutions cannot be counted as a marginal benefit of a hypothetical IMF Extended Fund Facility. Already secured resources are moved to the baseline; only incremental, liquid, timely and non-displaced financing is rewarded. The analysis combines corrected scenario scoring, poverty and inequality measures, maternal-child mortality indicators, public investment multipliers, self-defeating consolidation diagnostics, Monte Carlo uncertainty, leave-one-criterion-out tests and adverse execution assumptions. The results do not support an orthodox IMF-first strategy or front-loaded fiscal consolidation. The strongest designs are concessional multilateral packages with verified net additionality, productive execution, social-health floors and protection against poverty and maternal-child deterioration. An IMF arrangement is valuable only if it adds resources or credibility that were not already secured. The conclusion is that sustainability must be evaluated as a joint financial and social condition: closing cash gaps by weakening growth, poverty, inequality or maternal-child outcomes can be macroeconomically self-defeating.
\end{abstract}
\textbf{Keywords:} Bolivia; IMF; Extended Fund Facility; net financing additionality; fiscal consolidation; maternal mortality; infant mortality; poverty; inequality; fiscal multipliers; multilateral financing.
\tableofcontents
\newpage

\section{Introduction}
Bolivia's policy problem is not well described by a binary choice between accepting or rejecting an IMF Extended Fund Facility. A binary formulation hides the two margins that matter most. The first is the financial margin: whether the country obtains enough liquid, timely and concessional external financing to avoid an uncontrolled adjustment through reserves, imports, inflation, arrears and exchange-rate pressure. The second is the social margin: whether the stabilization path protects poverty, extreme poverty, inequality, maternal-child health and minimum public-service capacity while the fiscal and external accounts are being repaired. A program that closes the Treasury cash gap but raises poverty, weakens maternal-child survival, cuts complementary health spending or destroys the public investment pipeline may be financially tidy in the short run and macroeconomically self-defeating in the medium run.

The main correction in this paper is therefore methodological and substantive at the same time. Bolivia has already received, contracted or negotiated financing from multilateral institutions. Those flows cannot be treated as a new marginal benefit of a hypothetical IMF program. Financing already secured belongs in the baseline. Financing approved but not yet disbursed belongs in a pipeline with timing, liquidity and realization risk. Financing merely announced belongs in a more uncertain pipeline. An IMF arrangement should receive credit only for net additionality: additional liquid resources, accelerated disbursements, improved concessionality, maturity extension, reserve support, credibility effects that materially unlock new flows, or debt reprofiling that reduces near-term cash pressure. Without that correction, the model would double-count resources from institutions such as the IDB, CAF, the World Bank or FONPLATA and would mechanically overstate the value of an IMF-catalytic scenario.

The social correction is equally important. The analysis does not treat poverty, inequality and maternal-child health as residual variables to be checked after the fiscal model is solved. They enter the objective function and the constraints. This matters because consolidation can be internally contradictory. If current spending cuts fall on health operations, nutrition, transportation access, maintenance, social protection or administrative capacity required for project execution, the same consolidation that improves the fiscal balance can depress output, reduce revenue, raise poverty, increase preventable mortality risk and weaken political feasibility. The relevant criterion is not austerity versus spending, but the composition and timing of the adjustment.

The paper uses a corrected scenario database, public investment and social-health series, household-based poverty and inequality measures, diagnostic fiscal-multiplier estimates, social-health models with lags, stress tests, Monte Carlo weight uncertainty and leave-one-criterion-out checks. The results are not presented as causal identification. They are a structured diagnostic framework for comparing policy architectures under explicit assumptions. The central result is that an orthodox IMF-first program does not dominate. The leading designs are those that combine net additional financing, concessionality, execution capacity, high-multiplier investment and a verifiable maternal-child and poverty floor. If a sufficiently liquid and concessional multilateral package without the IMF is available, it can dominate an EFF. The IMF becomes attractive only if it adds something that was not already secured.

\begin{table}[!htbp]\centering\caption{Main data blocks and analytical roles.}\label{tab:sources}
\scriptsize
\begin{tabularx}{\textwidth}{@{}LLL@{}}
\toprule
Block & Main sources & Role \\
\midrule
Financial and external block & BCB/MEFP, external debt, cash-gap and balance-of-payments tables & Baseline liquidity, debt-service and external stress. \\
Investment block & Public investment by sector, department and funding source & Diagnostic multiplier and project-composition screening. \\
Household block & Survey-based poverty, extreme poverty, vulnerability and weighted Gini & Social floor and inequality constraints. \\
Social-health block & WDI/PIP/UN child-mortality and maternal-mortality indicators & Maternal-child risk, health floor and social sustainability. \\
Scenario block & Corrected scenario universe with net additionality, execution, delays and self-defeating consolidation & Ranking, stress tests, robustness and policy interpretation. \\
\bottomrule
\end{tabularx}
\end{table}

\section{Data, scenarios and modeling architecture}
The dataset integrates fiscal, external, investment, household and social-health information in a reproducible SQLite structure. The financial block includes cash-gap proxies, external public debt, external debt service, public-sector financing, balance-of-payments indicators and reserve stress variables. The investment block includes public investment by sector, department and funding source, plus diagnostic multiplier estimates based on local projections and lagged shares. The household block includes poverty, extreme poverty, vulnerability and a weighted Gini calculation using person weights from household expansion factors and household size. The social-health block adds maternal mortality, maternal deaths, neonatal mortality, infant mortality, under-five mortality, sanitation, health spending and related World Development Indicators.

The scenario universe is intentionally broader than a conventional IMF/no-IMF comparison. It includes orthodox EFF consolidation, productive EFF designs, catalytic EFF designs with verified additionality, already-committed multilateral financing with productive execution, incremental concessional multilateral financing without the IMF, selective consolidation with a health floor, debt reprofiling or debt swaps for social investment, progressive tax reform with social investment, subsidy reform with targeted transfers, state-owned-enterprise reform with reinvestment, gradual no-IMF adjustment, hard adjustment, excessive consolidation, forced domestic financing and no adjustment. This broader universe is necessary because a narrow choice set can make the preferred option look stronger than it really is.

Each scenario is scored through a vector of criteria. The first group measures financial sustainability: net additional financing, cash-gap pressure, external-gap pressure, debt risk and restructuring value. The second group measures social sustainability: poverty, extreme poverty, inequality, maternal-child risk, health floor and social protection. The third group measures productive viability: investment-health linkage, multiplier premium and execution feasibility. The fourth group penalizes self-defeating consolidation, defined as a configuration in which fiscal adjustment improves the short-run balance but damages output, poverty, revenue capacity, social stability or health outcomes enough to undermine the adjustment itself.

The integrated utility is computed as
\[
U_s = \sum_k w_k \tilde{x}_{s,k},
\]
where \(\tilde{x}_{s,k}\) denotes a normalized criterion for scenario \(s\), transformed so that higher values are always better, and \(w_k\) is the weight assigned to criterion \(k\). Net financing additionality is not the same as gross financing. It is defined as
\[
A_s = L_s\left[ P_s \rho_s + N_s(1-d_s)\eta_s \right] - B_s,
\]
where \(P_s\) is the financing pipeline, \(\rho_s\) is the incremental share of that pipeline, \(N_s\) is gross new financing, \(d_s\) is the displacement rate, \(\eta_s\) is expected realization, \(L_s\) is the liquidity share and \(B_s\) is the amount already secured and therefore moved to the baseline rather than rewarded as a new benefit. This expression is deliberately conservative: it prevents the scenario score from rising simply because a resource that Bolivia already had is relabeled as a marginal gain.

The self-defeating consolidation component is based on the interaction among adjustment intensity, selectivity, protected current spending, financing shortfall, execution capacity, poverty response and maternal-child risk. The logic is not that every fiscal consolidation is destructive. Selective consolidation can be necessary and welfare-improving when it removes low-multiplier current spending, regressive subsidies, inefficient transfers and quasi-fiscal losses while preserving social and operational spending. The danger is front-loaded consolidation without compensating investment, weak execution and no social floor.

\begin{table}[!htbp]\centering\caption{Financing audit in percent of GDP: secured flows are baseline, not marginal benefits.}\label{tab:additionality}
\scriptsize
\begin{tabularx}{\textwidth}{@{}LLLLL@{}}
\toprule
Scenario & Already secured & Pipeline & Net additional & Double-count prevented \\
\midrule
Orthodox EFF with accelerated consolidation & 1.40 & 0.40 & 1.32 & 1.76 \\
Incremental concessional multilateral package without the IMF & 2.40 & 2.50 & 1.01 & 3.77 \\
Catalytic multilateral EFF with verified net additionality & 2.20 & 1.60 & 0.76 & 3.24 \\
Hard adjustment & 1.40 & 0.30 & 0.75 & 1.68 \\
Already-committed multilateral package with productive execution and social floor & 2.80 & 2.20 & 0.72 & 4.12 \\
Productive EFF with maternal-child floor and net additionality & 1.60 & 0.70 & 0.71 & 2.16 \\
Consolidación fiscal prioritaria sin arquitectura social-productiva & 1.40 & 0.40 & 0.70 & 1.77 \\
Debt reprofiling or swap for verifiable social investment & 1.20 & 1.40 & 0.53 & 1.83 \\
\bottomrule
\end{tabularx}
\end{table}

\section{Financial sustainability and net additionality}
The corrected financial ranking changes the interpretation of the IMF question. Under the gross-financing view, a catalytic EFF appears naturally attractive because it is associated with multilateral resources. Under the net-additionality view, that logic is no longer sufficient. If IDB, CAF, World Bank or FONPLATA resources are already contracted, approved or expected independently of an IMF program, they are not an IMF benefit. The EFF must show that it accelerates, enlarges, cheapens, makes more liquid or de-risks financing beyond what Bolivia already had. This correction moves the best no-IMF multilateral packages upward and moves the IMF scenarios downward unless they deliver verified additionality.

The leading corrected result is not anti-IMF in a dogmatic sense. It is anti-double-counting. The best-ranked scenario is incremental concessional multilateral financing without the IMF, followed closely by the already-committed multilateral package with productive execution and a social floor. A catalytic EFF with verified net additionality remains competitive, but it no longer dominates. That is the central policy lesson: the IMF is useful only if it changes the feasible set. If it merely accompanies financing that would have arrived anyway, its conditionality and political costs may outweigh its marginal financial value.

\begin{table}[!htbp]\centering\caption{Corrected integrated ranking under net financing additionality.}\label{tab:ranking}
\scriptsize
\begin{tabularx}{\textwidth}{@{}LLLL@{}}
\toprule
Rank & Scenario & Family & Score \\
\midrule
1 & Incremental concessional multilateral package without the IMF & No-IMF multilateral & 0.724 \\
2 & Already-committed multilateral package with productive execution and social floor & No-IMF multilateral & 0.716 \\
3 & Catalytic multilateral EFF with verified net additionality & Catalytic EFF & 0.688 \\
4 & Productive EFF with maternal-child floor and net additionality & Productive EFF & 0.679 \\
5 & Selective multilateral consolidation with a health floor & Selective multilateral & 0.598 \\
6 & Debt reprofiling or swap for verifiable social investment & Social debt reprofiling & 0.595 \\
7 & Progressive tax reform with social investment & Progressive fiscal reform & 0.524 \\
8 & Subsidy reform with targeted transfers and investment & Subsidy reform & 0.515 \\
\bottomrule
\end{tabularx}
\end{table}

\begin{figure}[!htbp]
\centering
\includegraphics[width=0.96\textwidth]{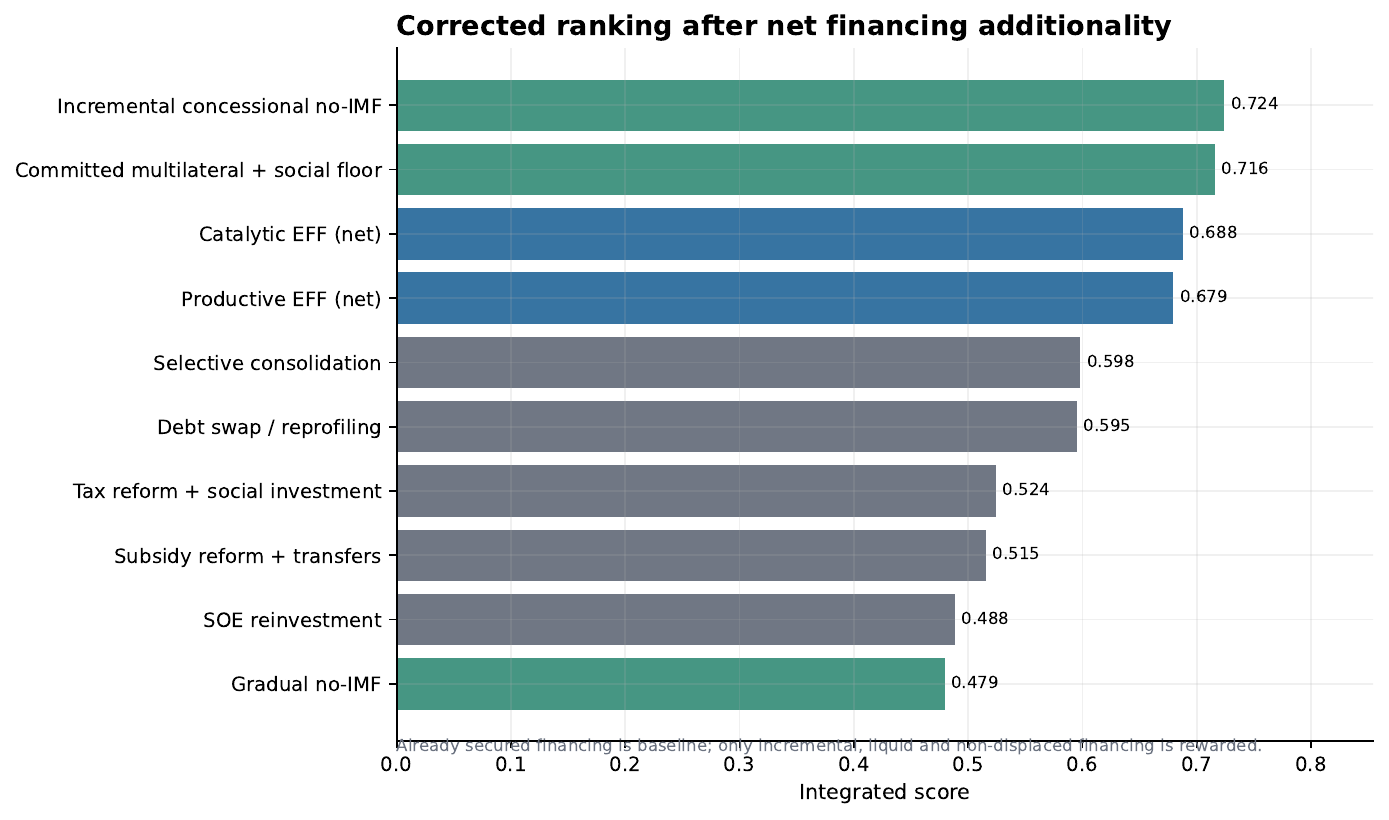}
\caption{Corrected integrated ranking after moving already secured multilateral financing to the baseline.}
\end{figure}

Monte Carlo uncertainty confirms the same ordering in probabilistic terms, although the interpretation must remain diagnostic. The probabilities are not objective forecasts of political success. They are frequencies generated by perturbing weights and measurement noise inside the model. Their value is comparative: they show whether a ranking depends on a single arbitrary weight vector or whether it survives broad preference variation. The leading no-IMF multilateral scenario wins most often because it combines net additionality, concessionality, low self-defeating risk and social protection without requiring an IMF label. The already-committed multilateral package is also robust because the model credits execution and social protection rather than gross financing.

\begin{table}[!htbp]\centering\caption{Monte Carlo ranking with weight uncertainty and measurement noise.}\label{tab:mc}
\scriptsize
\begin{tabularx}{\textwidth}{@{}LLLLLL@{}}
\toprule
MC rank & Scenario & Win prob. & Top-3 prob. & Mean rank & Mean score \\
\midrule
1 & Incremental concessional multilateral package without the IMF & 65.5\% & 99.3\% & 1.39 & 0.705 \\
2 & Already-committed multilateral package with productive execution and social floor & 32.2\% & 98.5\% & 1.79 & 0.694 \\
3 & Catalytic multilateral EFF with verified net additionality & 1.2\% & 65.9\% & 3.31 & 0.647 \\
4 & Productive EFF with maternal-child floor and net additionality & 1.1\% & 32.2\% & 3.71 & 0.635 \\
5 & Debt reprofiling or swap for verifiable social investment & 0.0\% & 2.9\% & 5.44 & 0.563 \\
6 & Orthodox EFF with accelerated consolidation & 0.0\% & 0.2\% & 10.96 & 0.404 \\
7 & Selective multilateral consolidation with a health floor & 0.0\% & 1.1\% & 5.47 & 0.564 \\
8 & Progressive tax reform with social investment & 0.0\% & 0.0\% & 7.69 & 0.487 \\
\bottomrule
\end{tabularx}
\end{table}

\begin{figure}[!htbp]
\centering
\includegraphics[width=0.96\textwidth]{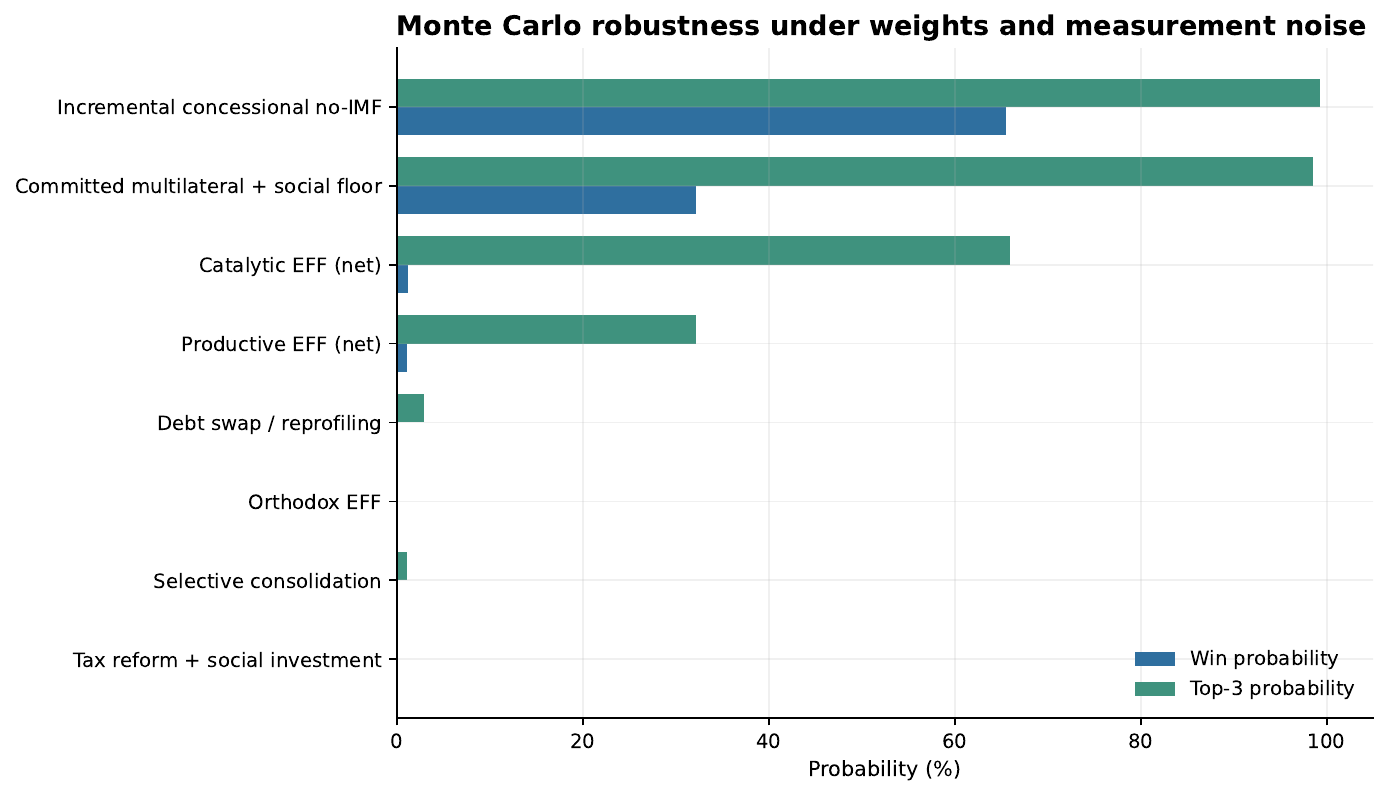}
\caption{Monte Carlo robustness under uncertain criterion weights and measurement noise.}
\end{figure}

\begin{figure}[!htbp]
\centering
\includegraphics[width=0.96\textwidth]{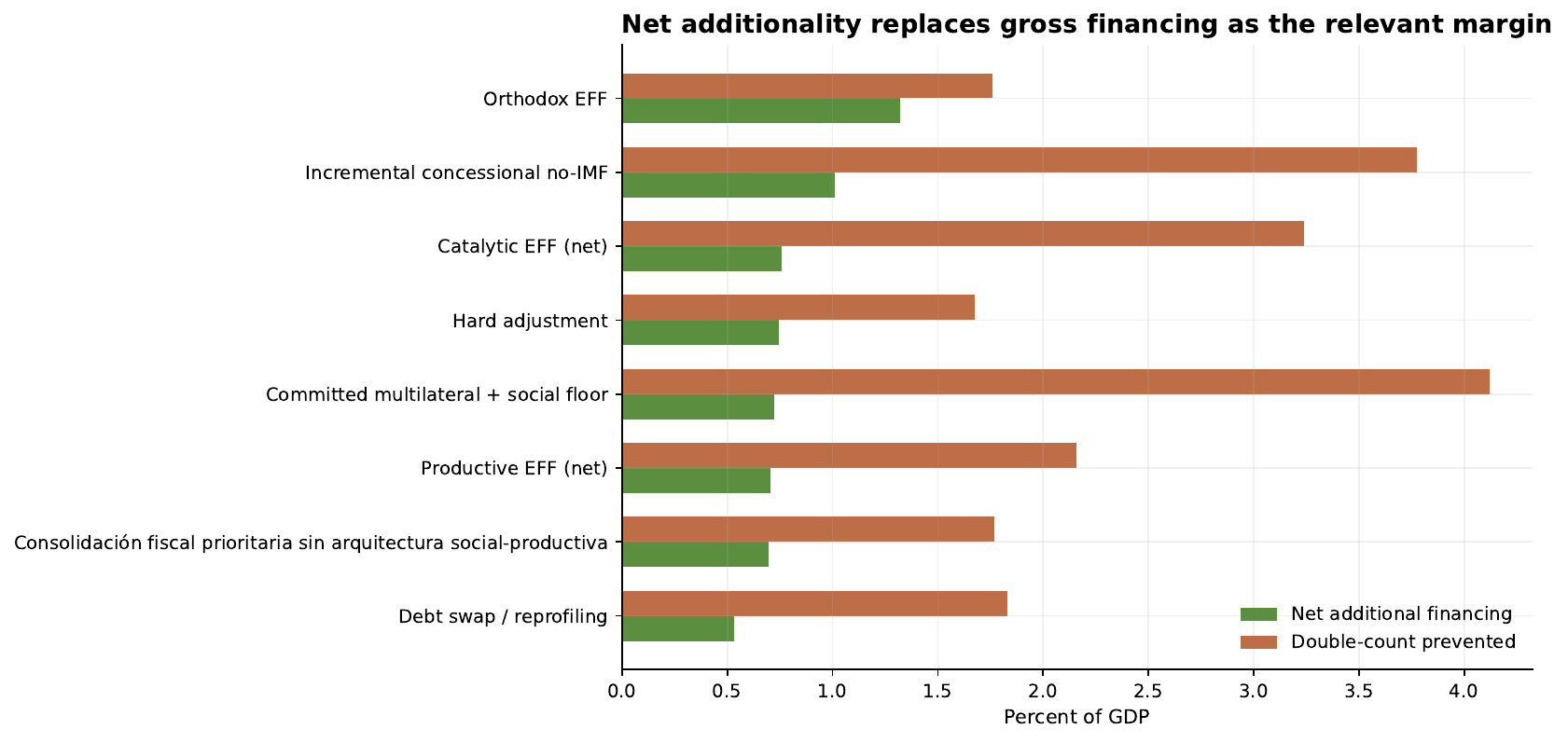}
\caption{Net additionality and double-count prevention. Already secured financing is treated as baseline, not as a marginal program gain.}
\end{figure}

\section{Verifiable social sustainability}
The social block is not an ethical appendix. It is a macroeconomic constraint. Bolivia's historical social-health trajectory shows large improvements in maternal-child outcomes, but the latest levels still make health floors relevant for any stabilization program. Maternal mortality declined from 518 deaths per 100,000 live births in 1985 to 146 in 2023. Infant mortality fell from 186.4 deaths per 1,000 live births in 1960 to 14.2 in 2024, while under-five mortality fell from 285.4 to 15.7 over the same broad period. Those improvements should not be interpreted as automatic future progress. A fiscal and external shock can affect access to health facilities, prenatal care, nutrition, sanitation, medicines, transportation and household income. The model therefore treats maternal-child risk as a protected criterion.

Poverty and inequality are included for the same reason. Fiscal consolidation that raises poverty or extreme poverty can become macroeconomically self-defeating even when the primary balance improves. The weighted 2024 household-survey Gini calculation is approximately 0.416, while national poverty remains high in the latest available data. These variables influence the ranking directly and also indirectly through social-health risk. The point is not that all poverty changes can be causally assigned to an EFF or a multilateral package. The point is that no stabilization strategy should be evaluated as successful if its financing plan relies on social deterioration that undermines health, demand and legitimacy.

\begin{table}[!htbp]\centering\caption{Poverty, inequality and maternal-child reference indicators.}\label{tab:health}
\scriptsize
\begin{tabularx}{\textwidth}{@{}LLLLLLL@{}}
\toprule
Indicator & First year & First value & Latest year & Latest value & Total change & Obs. \\
\midrule
Maternal mortality ratio & 1985 & 518.0 & 2023 & 146.0 & -71.8\% & 39 \\
Maternal deaths & 1985 & 1200.0 & 2023 & 380.0 & -68.3\% & 39 \\
Neonatal mortality & 1966 & 68.6 & 2024 & 7.3 & -89.4\% & 59 \\
Infant mortality & 1960 & 186.4 & 2024 & 14.2 & -92.4\% & 65 \\
Under-five mortality & 1960 & 285.4 & 2024 & 15.7 & -94.5\% & 65 \\
Gini index & 1990 & 42.0 & 2024 & 40.9 & -2.6\% & 27 \\
National poverty headcount & 2016 & 43.0 & 2024 & 37.7 & -12.3\% & 9 \\
Poverty at \$3.00/day & 1990 & 5.8 & 2024 & 2.1 & -63.8\% & 27 \\
Poverty at \$4.20/day & 1990 & 15.7 & 2024 & 4.9 & -68.8\% & 27 \\
Poverty at \$8.30/day & 1990 & 45.7 & 2024 & 17.4 & -61.9\% & 27 \\
Health expenditure (\% GDP) & 2000 & 4.4 & 2023 & 6.9 & 57.4\% & 24 \\
Basic sanitation access & 2000 & 34.9 & 2024 & 71.6 & 105.0\% & 25 \\
\bottomrule
\end{tabularx}
\end{table}

\begin{figure}[!htbp]
\centering
\includegraphics[width=0.9\textwidth]{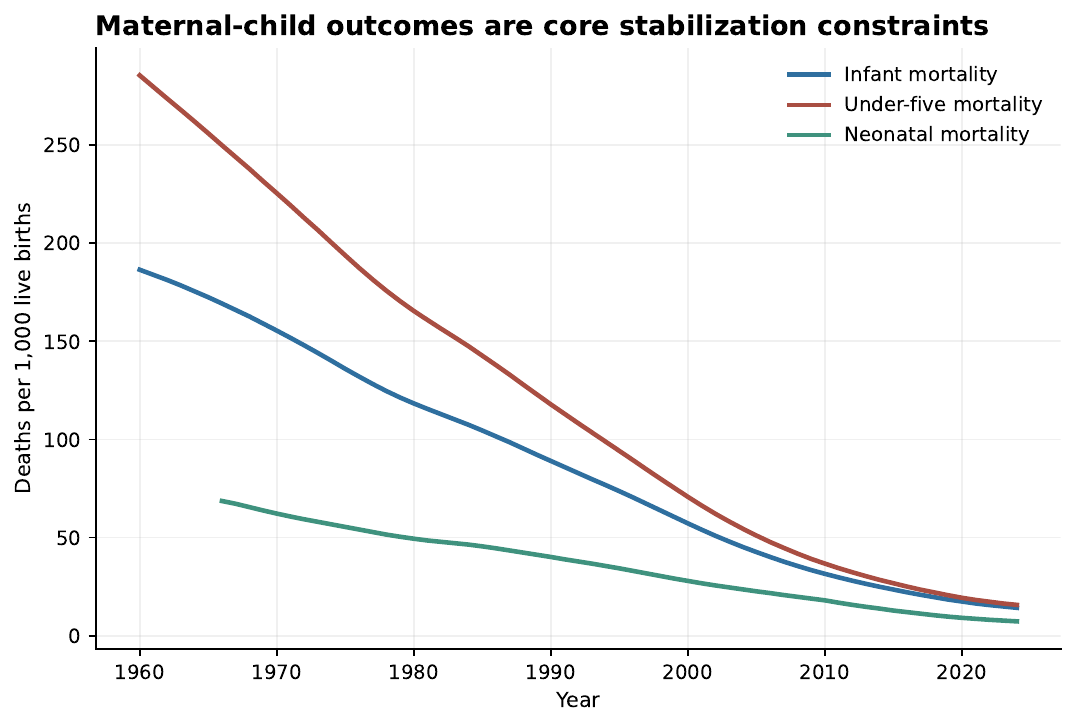}
\caption{Long-run child-mortality decline and the need to protect maternal-child floors during adjustment.}
\end{figure}

The social-health models connect changes in maternal and child mortality outcomes to a social-health risk index, inequality, growth, sanitation and health spending. The sample is small and annual; therefore, the models are diagnostic rather than causal. Their function is to discipline the ranking: health risk is not ignored, and scenarios that finance stabilization by weakening social capacity are penalized. The strongest lesson is that maternal-child outcomes respond to slow-moving structural variables and to access constraints, so the program's social floor must be operational, not rhetorical.

\begin{table}[!htbp]\centering\caption{OLS-HAC social-health diagnostic models.}\label{tab:models}
\scriptsize
\begin{tabularx}{\textwidth}{@{}LLLL@{}}
\toprule
Outcome model & Status & Obs. & Diagnostic \$R\textasciicircum{}2\$ \\
\midrule
Delta log maternal mortality & ok & 23 & 0.487 \\
Delta log neonatal mortality & ok & 24 & 0.584 \\
Delta log infant mortality & ok & 24 & 0.331 \\
Delta log under-five mortality & ok & 24 & 0.412 \\
\bottomrule
\end{tabularx}
\end{table}

\begin{table}[!htbp]\centering\caption{Largest social-health model coefficients by absolute test statistic.}\label{tab:coef}
\scriptsize
\begin{tabularx}{\textwidth}{@{}LLLLL@{}}
\toprule
Outcome & Term & Coef. & \$t\$ & p-value \\
\midrule
Neonatal mortality rate & social\_health\_risk\_index\_0\_100\_L1 & 0.000 & 5.55 & 0.000 \\
Infant mortality rate & social\_health\_risk\_index\_0\_100\_L0 & 0.000 & 3.85 & 0.000 \\
Under-five mortality rate & social\_health\_risk\_index\_0\_100\_L0 & 0.000 & 3.16 & 0.002 \\
Maternal mortality ratio & basic\_sanitation\_L0 & -0.032 & -2.91 & 0.004 \\
Maternal mortality ratio & social\_health\_risk\_index\_0\_100\_L0 & -0.014 & -2.49 & 0.013 \\
Maternal mortality ratio & gov\_health\_exp\_pc\_usd\_L0 & 0.011 & 1.96 & 0.050 \\
Maternal mortality ratio & current\_health\_exp\_pc\_usd\_L0 & -0.007 & -1.69 & 0.090 \\
Neonatal mortality rate & gdp\_growth\_L1 & -0.001 & -1.60 & 0.110 \\
Maternal mortality ratio & gini\_index\_L0 & 0.023 & 1.40 & 0.163 \\
Infant mortality rate & gdp\_growth\_L0 & -0.000 & -0.38 & 0.703 \\
Under-five mortality rate & gdp\_growth\_L0 & -0.000 & -0.20 & 0.844 \\
\bottomrule
\end{tabularx}
\end{table}

\begin{table}[!htbp]\centering\caption{Temporal validation of social-health prediction models.}\label{tab:cv}
\scriptsize
\begin{tabularx}{\textwidth}{@{}LLLLL@{}}
\toprule
Outcome & Model & Test obs. & RMSE & MAE \\
\midrule
Infant mortality rate & GradientBoosting & 48 & 0.020 & 0.015 \\
Infant mortality rate & ElasticNet & 48 & 0.025 & 0.021 \\
Maternal mortality ratio & Ridge & 15 & 0.166 & 0.086 \\
Maternal mortality ratio & GradientBoosting & 15 & 0.167 & 0.085 \\
Neonatal mortality rate & GradientBoosting & 44 & 0.017 & 0.013 \\
Neonatal mortality rate & Ridge & 44 & 0.025 & 0.017 \\
Under-five mortality rate & GradientBoosting & 48 & 0.020 & 0.015 \\
Under-five mortality rate & ElasticNet & 48 & 0.027 & 0.022 \\
\bottomrule
\end{tabularx}
\end{table}

\begin{figure}[!htbp]
\centering
\includegraphics[width=0.86\textwidth]{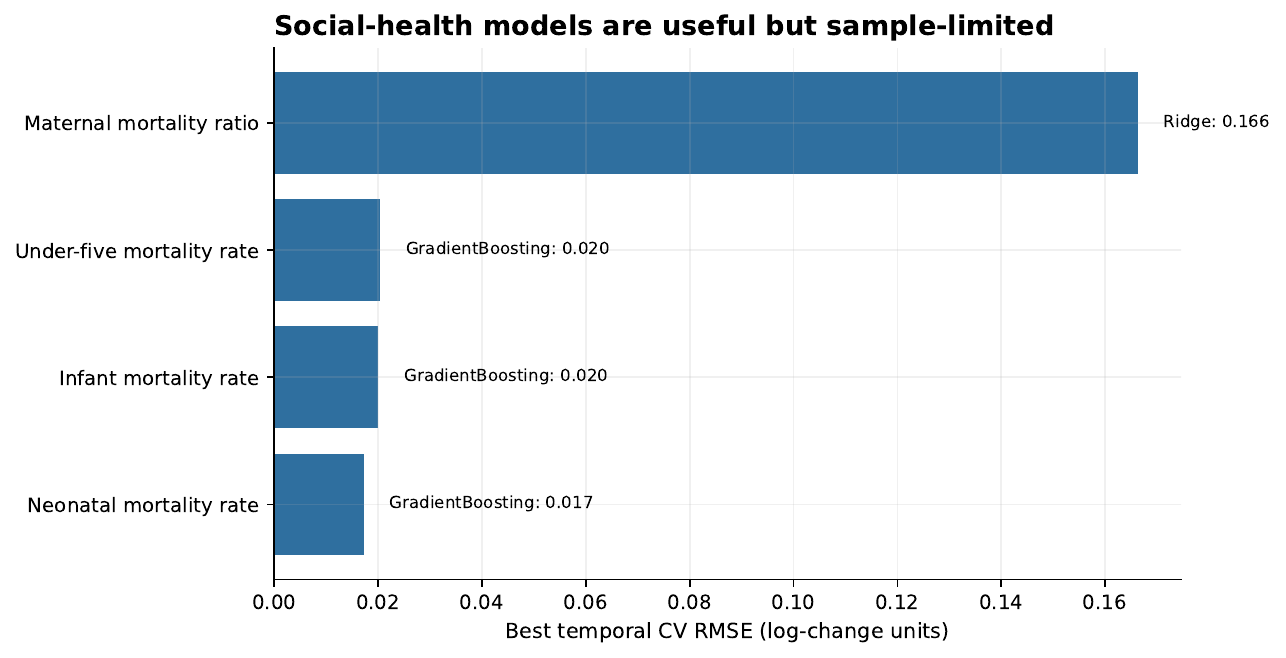}
\caption{Temporal validation of diagnostic social-health models. The exercise supports risk screening, not causal claims.}
\end{figure}

\section{Multipliers, investment composition and time-to-signal}
The investment component changes the interpretation of debt. Debt used to cover low-multiplier current spending worsens the numerator of the debt-to-GDP ratio without generating a meaningful denominator effect. Debt used to finance high-return public investment can improve the denominator, but only if execution is fast enough, leakages are controlled, import content does not absorb the impulse, and recurrent complementary spending is protected. This is why the model does not ask whether debt is good or bad in general. It asks whether the financing is additional, concessional and productively allocated.

The sectoral multiplier estimates are not structural causal parameters. They are diagnostic signals from historical associations between public investment composition and sectoral output dynamics. High estimates can reflect real spillovers, but also simultaneity, omitted variables, small investment bases or anticipation effects. For that reason, the paper caps and discounts multipliers before using them in scenario scoring. The most useful interpretation is ordinal rather than literal: transport, sanitation, education-related investment, water and selected productive infrastructure are stronger candidates than undifferentiated current expenditure.

\begin{table}[!htbp]\centering\caption{Diagnostic fiscal-multiplier ranking by public-investment sector.}\label{tab:mult}
\scriptsize
\begin{tabularx}{\textwidth}{@{}LLLLLLL@{}}
\toprule
Rank & Sector & \$h=1\$ & \$h=2\$ & \$h=3\$ & Robust score & 2024 investment \\
\midrule
1 & Education and culture & 9.97 & 11.51 & 10.68 & 10.07 & 198.5 \\
2 & Transport & 4.54 & 5.97 & 6.98 & 9.53 & 572.0 \\
3 & Basic sanitation & 6.82 & 7.94 & 4.87 & 7.99 & 144.1 \\
4 & Hydrocarbons & 2.46 & 6.40 & 9.75 & 6.39 & 260.9 \\
5 & Industry and tourism & 16.66 & 20.47 & 16.79 & 6.20 & 231.8 \\
6 & Mining & 14.12 & 9.76 & 13.17 & 5.33 & 78.2 \\
7 & Urban development and housing & 4.55 & 6.22 & 8.57 & 5.18 & 244.8 \\
8 & Water resources & 9.61 & 16.24 & 32.19 & 4.89 & 16.2 \\
9 & Energy & 3.66 & 5.45 & 5.10 & 4.00 & 204.3 \\
10 & Agriculture & 6.60 & 6.00 & 6.06 & 3.82 & 160.5 \\
\bottomrule
\end{tabularx}
\end{table}

\begin{table}[!htbp]\centering\caption{Illustrative high-multiplier allocation after leakage and delay discount.}\label{tab:alloc}
\scriptsize
\begin{tabularx}{\textwidth}{@{}LLLLL@{}}
\toprule
Sector & Share & Allocation & Conservative multiplier & Expected GDP gain \\
\midrule
Education and culture & 30\% & 425.8 & 3.60 & 1532.8 \\
Transport & 25\% & 354.8 & 2.69 & 953.3 \\
Basic sanitation & 20\% & 283.9 & 3.57 & 1014.4 \\
Hydrocarbons & 15\% & 212.9 & 2.88 & 613.2 \\
Industry and tourism & 10\% & 141.9 & 3.60 & 510.9 \\
\bottomrule
\end{tabularx}
\end{table}

\begin{figure}[!htbp]
\centering
\includegraphics[width=0.96\textwidth]{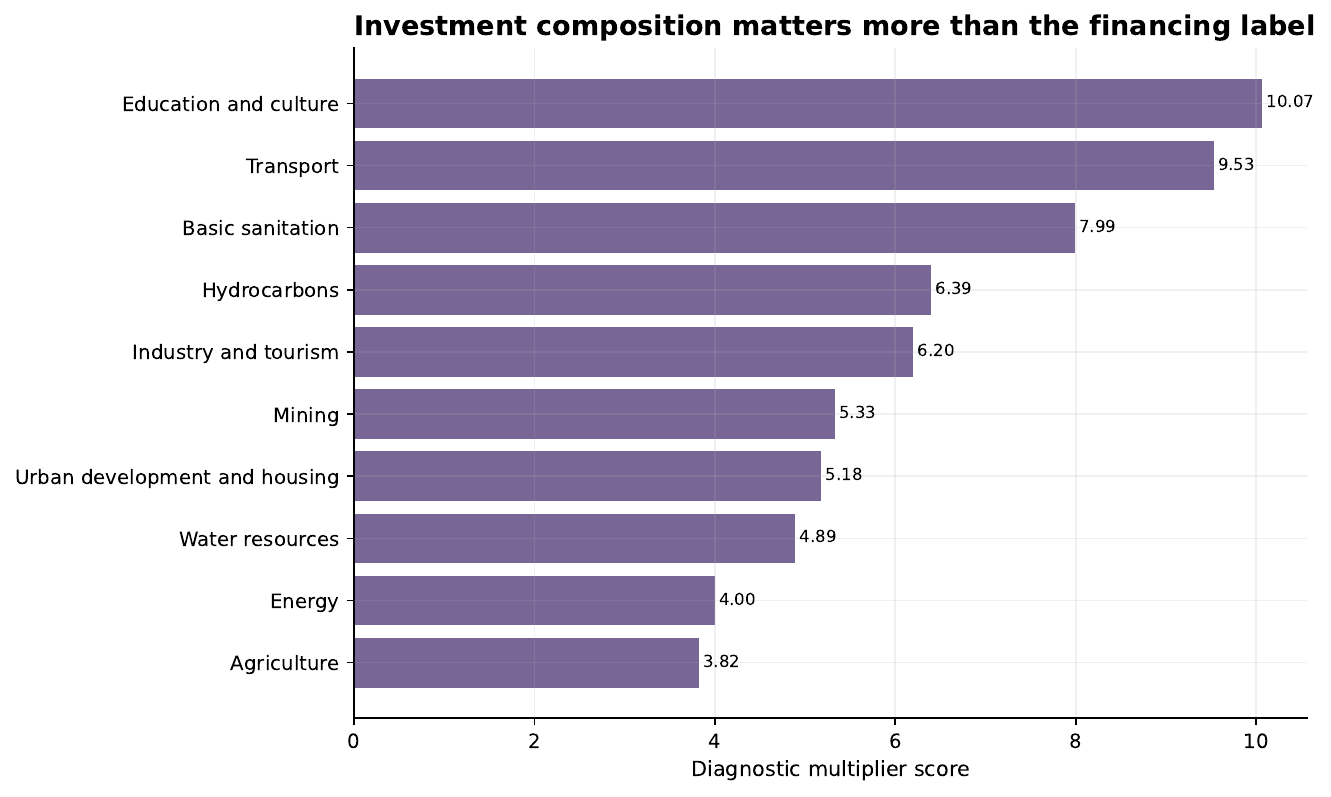}
\caption{Diagnostic multiplier priorities by sector. The estimates are used for screening and composition, not for causal certainty.}
\end{figure}

Time-to-signal is crucial. A productive package may be superior over a multi-year horizon, but it is not an instant stabilizer. The first year can still be contractionary if consolidation starts before investment reaches physical execution. This is the central danger in front-loaded adjustment: the macro model may show future gains while households experience immediate losses. The best policy architecture must therefore combine a financing bridge, phased consolidation, protected social spending and a project pipeline mature enough to spend without waste.

\begin{figure}[!htbp]
\centering
\includegraphics[width=0.96\textwidth]{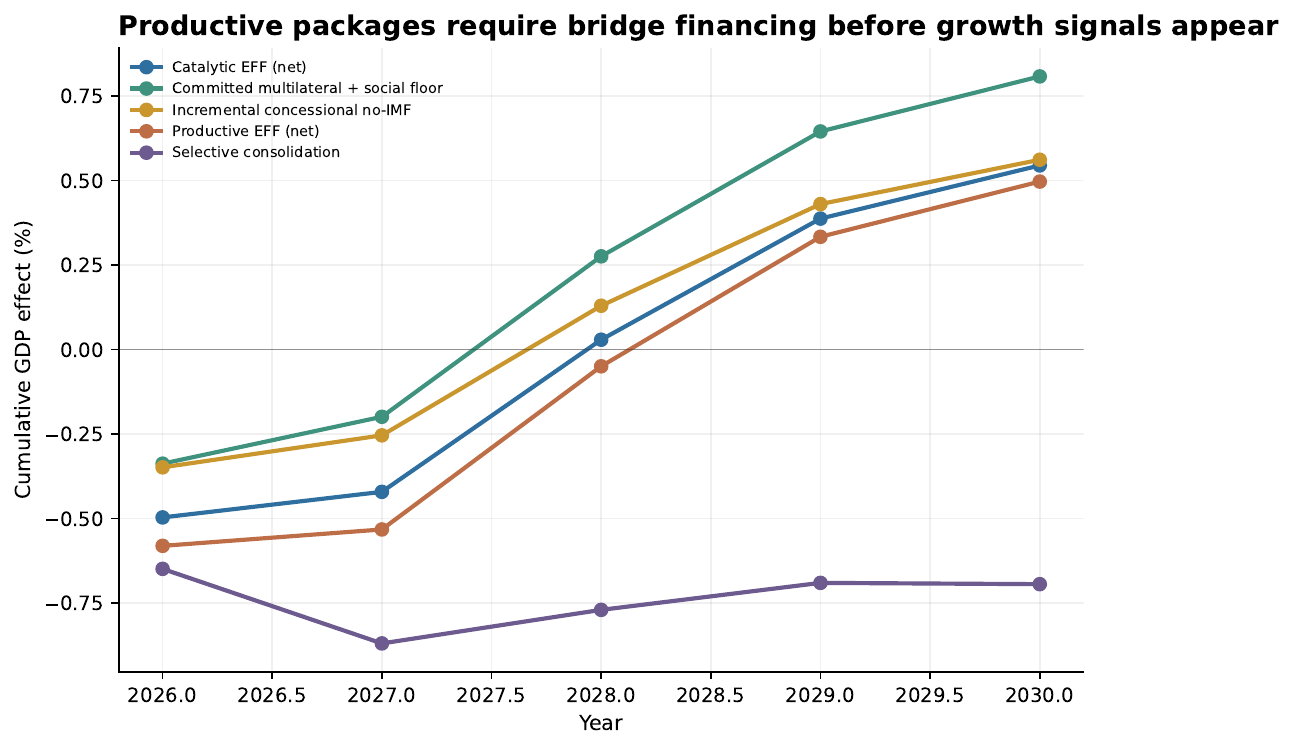}
\caption{Cumulative GDP-effect paths for leading scenarios. Productive packages need bridge financing before growth effects become visible.}
\end{figure}

\begin{table}[!htbp]\centering\caption{Macro-diagnostic model comparison under time-series cross-validation.}\label{tab:macro}
\scriptsize
\begin{tabularx}{\textwidth}{@{}LLLLLL@{}}
\toprule
Rank & Model & Obs. & CV RMSE & CV MAE & CV \$R\textasciicircum{}2\$ \\
\midrule
1 & GradientBoosting & 34 & 2.61 & 1.39 & 0.179 \\
2 & SVR\_rbf & 34 & 2.68 & 1.44 & 0.134 \\
3 & KNN & 34 & 2.80 & 1.59 & 0.052 \\
4 & ExtraTrees & 34 & 2.83 & 1.55 & 0.032 \\
5 & BayesianRidge & 34 & 2.85 & 1.64 & 0.021 \\
6 & RandomForest & 34 & 2.86 & 1.56 & 0.011 \\
7 & Ridge & 34 & 3.15 & 2.21 & -0.198 \\
8 & ElasticNet & 34 & 3.32 & 2.38 & -0.328 \\
\bottomrule
\end{tabularx}
\end{table}

\section{Robustness, adverse cases and policy interpretation}
A scenario is not convincing if it wins only under its own ideal assumptions. The robustness layer therefore subjects the ranking to alternative normative profiles, weight uncertainty, leave-one-criterion-out checks and self-defeating consolidation diagnostics. The corrected result is more credible than the previous gross-financing result because the leading no-IMF multilateral scenarios remain strong after double-count correction. The IMF scenarios do not disappear; they become conditional. They are valuable if they provide verified net additionality, accelerate disbursements, extend maturities, lower financing costs or anchor other resources that would not otherwise materialize.

The leave-one-criterion-out test shows whether the ranking is driven by one variable. The leading scenarios remain near the top when individual criteria are removed, but their interpretation changes. If net additionality is removed, scenarios with strong execution and social protection become closer. If the maternal-child floor receives less weight, financially hard scenarios look better, but only by ignoring the social failure channel. If execution feasibility is removed, high-multiplier packages are overvalued. This confirms that the ranking is not simply a financing table; it is an integrated sustainability screen.

\begin{table}[!htbp]\centering\caption{Rank of leading scenarios under alternative normative profiles.}\label{tab:profiles}
\tiny
\begin{tabularx}{\textwidth}{@{}LLLLLLLL@{}}
\toprule
Scenario & Anti-self-defeating consolidation & Balanced integrated & Conservative execution & Productive investment & Social-health & Strict financial & preferencia\_sin\_fmi \\
\midrule
Already-committed multilateral package with productive execution and social floor & 2 & 2 & 2 & 1 & 2 & 2 & 2 \\
Catalytic multilateral EFF with verified net additionality & 3 & 3 & 3 & 4 & 3 & 3 & 3 \\
Debt reprofiling or swap for verifiable social investment & 6 & 6 & 6 & 5 & 6 & 5 & 6 \\
Incremental concessional multilateral package without the IMF & 1 & 1 & 1 & 2 & 1 & 1 & 1 \\
Productive EFF with maternal-child floor and net additionality & 4 & 4 & 4 & 3 & 4 & 4 & 4 \\
Selective multilateral consolidation with a health floor & 5 & 5 & 5 & 6 & 5 & 6 & 5 \\
\bottomrule
\end{tabularx}
\end{table}

\begin{table}[!htbp]\centering\caption{Leave-one-criterion-out stability for leading scenarios.}\label{tab:loo}
\scriptsize
\begin{tabularx}{\textwidth}{@{}LLLL@{}}
\toprule
Scenario & Best rank & Worst rank & Mean rank \\
\midrule
Incremental concessional multilateral package without the IMF & 1 & 1 & 1.00 \\
Already-committed multilateral package with productive execution and social floor & 2 & 2 & 2.00 \\
Catalytic multilateral EFF with verified net additionality & 3 & 4 & 3.07 \\
Productive EFF with maternal-child floor and net additionality & 3 & 4 & 3.93 \\
Selective multilateral consolidation with a health floor & 5 & 6 & 5.20 \\
Debt reprofiling or swap for verifiable social investment & 5 & 6 & 5.80 \\
Progressive tax reform with social investment & 7 & 7 & 7.00 \\
Subsidy reform with targeted transfers and investment & 8 & 8 & 8.00 \\
\bottomrule
\end{tabularx}
\end{table}

\begin{figure}[!htbp]
\centering
\includegraphics[width=0.96\textwidth]{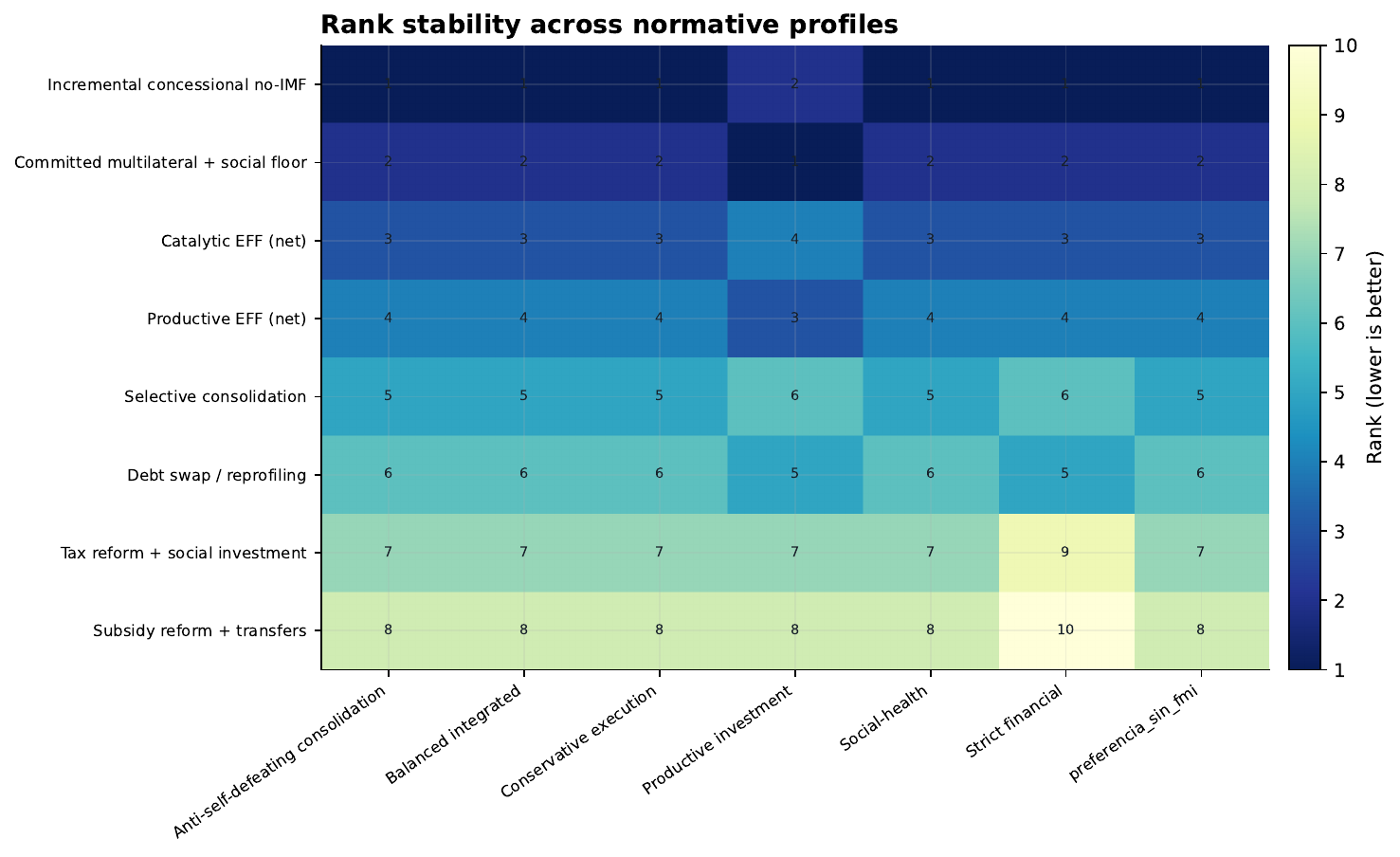}
\caption{Rank stability across profiles: financial, social-health, productive, execution-conservative and anti-self-defeating consolidation criteria.}
\end{figure}

\begin{figure}[!htbp]
\centering
\includegraphics[width=0.82\textwidth]{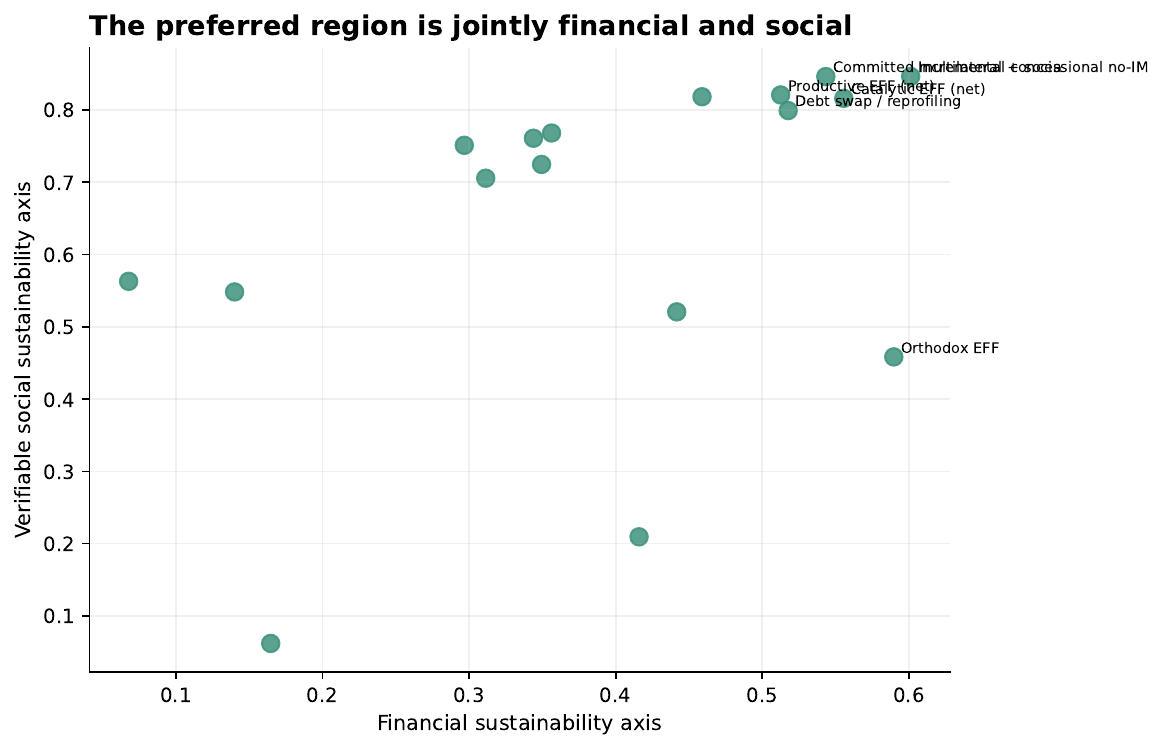}
\caption{Financial and social sustainability must be jointly satisfied.}
\end{figure}

The self-defeating consolidation results are the strongest warning against a mechanically orthodox program. Fiscal consolidation can be necessary, but its composition determines its macroeconomic sign. A selective package that removes low-return current spending and protects health, nutrition, social transfers, maintenance and execution capacity can support stabilization. A blunt package can reduce expenditure today and weaken revenue tomorrow. In the model, high self-defeating risk appears precisely when financing is insufficient, current cuts are large, selectivity is low, investment execution is weak, social protection is thin and poverty risk rises.

\begin{table}[!htbp]\centering\caption{Self-defeating consolidation diagnostics.}\label{tab:risk}
\scriptsize
\begin{tabularx}{\textwidth}{@{}LLLLLL@{}}
\toprule
Scenario & Shortfall & Self-defeat prob. & Poverty change & Maternal-child risk & First signal \\
\midrule
Subsidy reform with targeted transfers and investment & 1.84 & 100.0\% & 0.63 & 52.1 & 2030 \\
Hard adjustment & 1.08 & 100.0\% & 2.86 & 95.2 &  \\
State-owned-enterprise reform with productive reinvestment & 1.93 & 100.0\% & 0.68 & 55.0 & 2030 \\
Progressive tax reform with social investment & 1.71 & 100.0\% & 0.58 & 49.7 & 2029 \\
Forced domestic financing & 2.14 & 100.0\% & 1.06 & 72.6 &  \\
No-IMF gradual adjustment with partial financing & 1.68 & 100.0\% & 0.60 & 51.4 &  \\
Consolidación fiscal prioritaria sin arquitectura social-productiva & 1.12 & 100.0\% & 1.21 & 66.0 &  \\
Orthodox EFF with accelerated consolidation & 0.46 & 100.0\% & 1.49 & 65.8 &  \\
\bottomrule
\end{tabularx}
\end{table}

\begin{figure}[!htbp]
\centering
\includegraphics[width=0.96\textwidth]{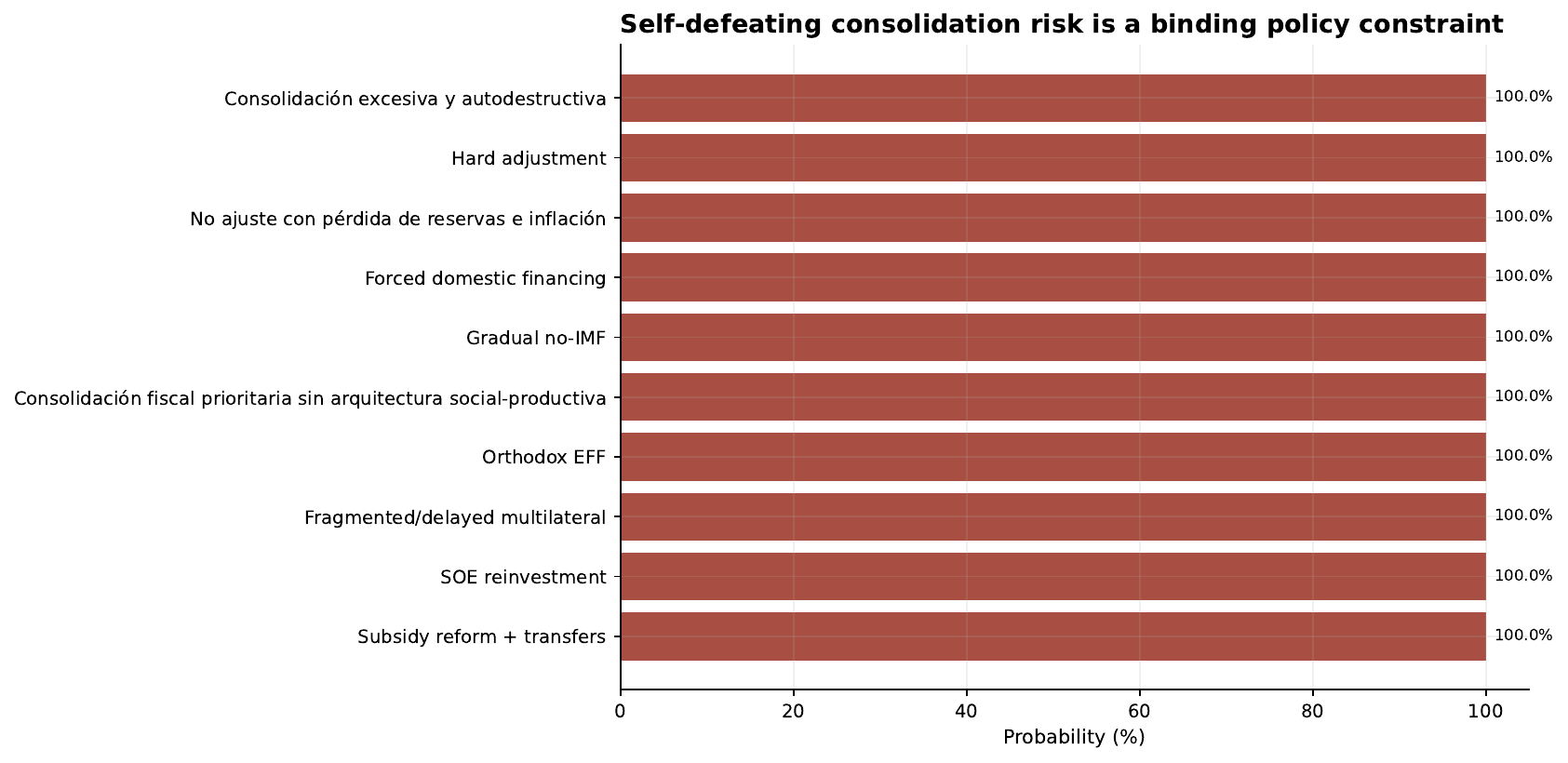}
\caption{Self-defeating consolidation risk under the corrected scenario universe.}
\end{figure}

The policy conclusion is conditional and deliberately narrow. If Bolivia can secure enough incremental concessional multilateral financing without the IMF, with timely liquidity, credible execution and a social-health floor, that option can dominate. If the already committed multilateral pipeline is large enough and can be converted into productive execution without deteriorating poverty or health, it is also highly competitive. If those conditions are not met, an EFF may still be useful, but only if it adds something net: new liquidity, stronger disbursement certainty, maturity extension, reserve support or catalytic effects that are demonstrably additional. An IMF program that mainly imposes front-loaded consolidation without additional financing, investment protection or a verifiable maternal-child floor is not attractive in this framework.

\section{Conclusion}
The corrected analysis changes the question. The relevant issue is not whether Bolivia should take an IMF program in the abstract. The relevant issue is whether any financing architecture closes the external and fiscal gap without destroying social sustainability and whether its financing is truly additional. Once already secured multilateral resources are placed in the baseline, the best-ranked scenarios are not orthodox IMF scenarios. They are multilateral, concessional, productively executed and socially protected designs. The IMF is potentially useful as a catalytic instrument, but only if it generates verified net additionality.

The most robust strategy is therefore a composition rule: secure liquid and concessional financing, prevent double counting, reduce current spending only where it is low-multiplier and non-essential, protect health and social floors, invest in high-multiplier sectors with mature projects, audit execution, and track poverty, inequality and maternal-child outcomes as core performance variables. Financial sustainability without social sustainability is unstable. Social sustainability without financing is fragile. The feasible path is the joint one: net additional financing plus verifiable social protection plus productive investment.

\end{document}